\documentclass[12pt]{article}
\usepackage{amsfonts,amsmath}
\usepackage{epsfig,graphics}

\renewcommand{\appendix}{%
\renewcommand{\section}{%
\newpage
\thispagestyle{plain}%
\secdef\Appendix\sAppendix}%
\setcounter{section}{0}%
\renewcommand{\thesection}{\Alph{section}}%
}
\newcommand{\Appendix}[2][?]{%
\refstepcounter{section}%
\addcontentsline{toc}{Addendum}%
{\protect\numberline{\appendixname~\thesection}#1}%
{\flushleft\LARGE\bfseries\appendixname\ \thesection\par
\centering#2\par}%
\sectionmark{#1}\vspace{\baselineskip}}

\newcommand{\sAppendix}[1]{%
{\flushright\large\bfseries\appendixname\par
\centering#1\par}%
\vspace{\baselineskip}}

\def\be{\begin{equation}}

\def\bea{\begin{eqnarray}}

\def\eea{\end{eqnarray}}

\newtheorem{definition}{Definition}[section]

\newtheorem{remark}[definition]{Remark}

\normalbaselines \oddsidemargin 0.5cm \evensidemargin 0.5cm
\begin{document}

\pagestyle{empty}

\rightline{}

 LAPTH-1215/07
\vskip 2cm

\begin{center}

{\Huge {\textbf{New connection formulae for some $q$-orthogonal
polynomials in $q$-Askey scheme}}}

\vspace{6mm}

{\bf \large  A. Yanallah \footnote{E-mail address~:
yanallahabdelkader@hotmail.com} and M. B. Zahaf \footnote{E-mail
address~: m\_b\_zahaf@yahoo.fr}}\setcounter{footnote}{0}\footnote{Visiting scientists at
LAPTH, Universit\'e de Savoie, CNRS, BP 110, F-74941
Annecy-le-Vieux, Cedex, France.}

\vspace{6mm}

\setcounter{footnote}{3}
\footnote{Laboratoire de recherche agr\'e\'e par le MESRS dans le
cadre du fond national de la recherche et du d\'eveloppement
technologique.} Laboratoire de Physique Quantique de la Mati\`ere et
de Mod\'elisations Math\'ematiques, Centre Universitaire de Mascara,
29000-Mascara, Alg\'erie\\

\end{center}
\vspace{4cm}
\begin{abstract}
{\small \noindent New nonlinear connection formulae of the
$q$-orthogonal polynomials, such continuous $q$-Laguerre, continuous
big $q$-Hermite, $q$-Meixner-Pollaczek and $q$-Gegenbauer
polynomials, in terms of their respective classical analogues are
obtained using a special realization of the $q$-exponential function
as infinite multiplicative series of ordinary exponential function.}
\end{abstract}

\pagestyle{plain} \setcounter{page}{1}


\newpage

\section{Introduction and motivation}
\setcounter{equation}{0}

In modern mathematical physics, hypergeometric and
$q$-hypergeometric functions have found their applications in the
development of the theory of difference equations and in quantum and
non commutative geometry. And in many results, like the theory of
lattice integrable models, Bethe ansatz and Toda systems
\cite{FR92}-\cite{Z92} for instance, they
 are formulated or realized in connection with these types of
mathematical functions. In this context and to illustrate a physical
application, we cite ref.\cite{MV93} where a representation of
$q$-hypergeometric functions of one variable was found in terms of
correlators of vertex operators made out of free scalar fields
propagating on Riemann sphere. Among these basic functions or
$q$-functions, there are polynomials which are structured in
schemes. An interesting one, so-called Askey-scheme \cite{KS} of
hypergeometric orthogonal polynomials, consists of all known sets of
orthogonal polynomials which can be defined in terms of a
hypergeometric function and their interrelations. The
$q$-Askey-scheme is the quantum version of the former, however the
hypergeometric orthogonal polynomials may admit several
$q$-analogues. Only few of these $q$-orthogonal polynomials possess
generating functions written in terms of $q$-exponential functions.
It is for this fact that  we deal with these $q$-polynomials in this
paper. Our interest here was motivated by the results of reference
\cite{Q} where it was established that the two Jackson's
$q$-exponentials
\begin{equation}
  e_q(z)=\sum_{k\in
\mathbb{N}}\frac{1}{(q;q)_k}z^k=\frac{1}{(z;q)_{\infty}},\qquad
  E_q(z)=\sum_{k\in
\mathbb{N}}\frac{q^{k(k-1)/2}}{(q;q)_k}z^k=(-z;q)_{\infty},
\end{equation}with $(a;q)_0=1$, $(a;q)_k=\prod_{j=0}^{k-1}(1-aq^j)$,
and $(a;q)_\infty=\prod_{j=0}^{\infty}(1-aq^j)$, could be expressed
respectively as the exponential of series as follows

\begin{equation}\label{quesne1}
  e_q(z)=\exp\left(\sum_{k\in
 \mathbb{N}^{*}}\frac{z^k}{k(1-q^k)}\right)
\end{equation}
and
\begin{equation}\label{quesne2}
  E_q(z)=\exp\left(\sum_{k\in
\mathbb{N}^{*}}\frac{(-1)^{k+1}z^k}{k(1-q^k)}\right)
\end{equation}
Furthermore, in the reference \cite {CJN}, the multiplicative series
form of the $q$-exponential were exploited to derive a new nonlinear
connection formula between $q$-orthogonal polynomials and their
classical versions, namely $q$-Hermite, $q$-Laguerre and
$q$-Gegenbauer polynomials. Their results are expressed in compact
form and some explicit examples are given. Also, the authors of
\cite{CJN} emphasized the possibility to extend their work for other
$q$-orthogonal polynomials such as little $q$-Jacobi ones. In the
present work we will take benefit of their idea to compute the
connection formula between other $q$-orthogonal polynomials,
appearing in the $q$-Askey scheme \cite{KS}, and their classical
counterparts, namely the continuous $q$-Laguerre, the continuous big
$q$-Hermite and the $q$-Meixner-Pollaczek polynomials and we give an
alternative
  connection
formula for the $q$-Gegenbauer polynomials distinct from the one
given in \cite{CJN}. In our knowledge, these cases have not been treated
before.  To proceed, similarly to the work of Chakrabarti
 {\em
et al} \cite{CJN}, we first give the generating functions of any
$q$-polynomials cited above and use, on one side, the series
development. On the other side, the Quesne formulae allow us to
express the $q$-exponential function as a product  series of the
classical exponential function. And this leads finally to the
connection formula, up to the resolution of a Diophantine partition
equation appearing during our computation for any examined cases.

As all the $q$-polynomials constituting the $q$-Askey scheme can be
defined in terms of the basic hypergeometric series ${}_r\phi_s$, we
recall here their expression (see for example \cite{GR}):
\begin{equation}
{}_r\phi_s(a_1,a_2,...,a_r;b_1,b_2,...,b_s;q;z)=
\sum_{n=0}^{\infty}\frac{(a_1;q)_n(a_2;q)_n...(a_r;q)_n}{(q;q)_n(b_1;q)_n(b_2;q)_n...(b_s;q)_n}\left[
(-1)^nq^{\frac{n(n-1)}{2}}\right]^{1+s-r}z^n
\end{equation}with
$q\neq0$. The ratio test shows that for generic values of the
parameters the radius of convergence is $\infty$, $1$ or $0$ for
$r<s+1$, $r=s+1$ or $r>s+1$ respectively. Since $(q^{-n};q)_k=0$ for
$k=n+1, n+2,...$, the series ${}_r\phi_s$ terminates if one of the
numerator parameters $\{a_i\}$ is of the form $q^{-n}$ with
$n=0,1,2,...$ and $q\neq 0$. The ${}_r\phi_s$ function is the
$q$-analogue of the hypergeometric function defined by
\begin{equation}
{}_rF_s(a_1,a_2,...,a_r;b_1,b_2,...,b_s;z)=\sum_{n=0}^{\infty}\frac{(a_1)_n(a_2)_n...(a_r)_n}{(b_1)_n(b_2)_n...(b_s)_n}
\frac{z^n}{n!}\end{equation} where $(a)_n$ denotes the Pochhammer
symbol defined by $$ (a)_0=1, \textrm{and} ~~
(a)_k=a(a+1)(a+2)...(a+k-1), k=1,2,...$$ When one of the numerator
parameters $a_i$ equals $-n$ where $n$ is a nonnegative integer this
hypergeometric series is a polynomial in $z$. Otherwise the radius
of convergence is $\infty$, $1$ or $0$ for $r<s+1$, $r=s+1$ or
$r>s+1$ respectively.
  \vskip 0.5cm

\section{Continuous $q$-Laguerre polynomials}
\setcounter{equation}{0}

The continuous $q$-Laguerre polynomials had manifested their
apparition in the rational solutions of the $q$-analogue of
Painlev\'e V differential equation \cite{masuda}, namely as the
entries of its associated determinant.
 They are defined by: \cite{KS}
\begin{eqnarray}
&&    P_n^{\alpha}(x|q)=\frac{(q^{\alpha+1};q)_n}{(q;q)_{n}}
{}_3\phi_2\left(q^{-n} ,q^{\frac{1}{ 2} \alpha+ \frac{1}{
4}}e^{i\theta} ,q^{\frac{1}{ 2} \alpha+ \frac{1}{
4}}e^{-i\theta};q^{\alpha+1},0;q;q
\right),\quad x=\cos\theta\\
&&=\frac{(q^{\frac{1}{ 2} \alpha+ \frac{3}{
4}}e^{-i\theta};q)_n}{(q;q)_{n}}q^{({\frac{1}{ 2} \alpha+ \frac{1}{
4}})n}e^{in\theta} {}_2\phi_1\left(q^{-n} ,q^{\frac{1}{ 2} \alpha+
\frac{1}{ 4}}e^{i\theta} ;q^{-{\frac{1}{ 2} \alpha+ \frac{1}{
4}}-n};q;q^{-{\frac{1}{ 2} \alpha+ \frac{1}{ 4}}}e^{-i\theta}\right)
\end{eqnarray}
 The
generating function of the continuous $q$-Laguerre polynomials is
given by \cite{KS}
\begin{eqnarray}\label {genrating1}
&&\textsf{G}_q^{\alpha}(x;t)\equiv \frac{(q^{\alpha+\frac{1}{2}}t;q)_\infty(q^{\alpha+1}t;q)_\infty}{(q^{\frac{1}{2}\alpha+\frac{1}{4}}e^{i\theta}t;q)_\infty(q^{\frac{1}{2}\alpha+\frac{1}{4}}e^{-i\theta}t;q)_\infty}\nonumber\\
&&=E_q(-q^{\alpha+\frac{1}{2}}t)E_q(-q^{\alpha+1}t)e_q(q^{\frac{1}{2}\alpha+\frac{1}{4}}e^{i\theta}t)e_q(q^{\frac{1}{2}\alpha+\frac{1}{4}}e^{-i\theta}t)
=\sum_{n\geq 0}P_n^{\alpha}(x|q)t^n.
\end{eqnarray}
In the $q\rightarrow 1$ limit, when $x$ is replaced by
$q^{\frac{x}{2}}$ in the above function (\ref{genrating1}), we find
the generating function
\begin{equation}\label {genrating2}
  \textsf{G}^{\alpha}(x,t)\equiv
  (1-t)^{-\alpha-1}\exp\left(\frac{xt}{t-1}\right)=\sum_{n=0}^{\infty}L_n^{\alpha}(x)t^n
\end{equation}for the classical
Laguerre polynomials \cite{AAR}
\begin{equation}
L_n^{\alpha}(x)=\frac{(\alpha+1)_n}{n!}{}_1F_1\left(-n;\alpha+1;x\right).
\end{equation}
 Using (\ref{quesne1}) and (\ref{quesne2}) the
left hand side of (\ref {genrating1}) is reformulated as
\begin{equation}\label{genrating4}
\textsf{G}_q^{\alpha}(x;t)=\prod_{k\in\mathbb{N}^*}\exp\left(\frac{-q^{(\alpha+\frac{1}{2})k}-q^{(\alpha+1)k}+2q^{(\frac{1}{2}\alpha+\frac{1}{4})k}\cos
k\theta}{k(1-q^k)}t^{k}\right).
\end{equation}
To express the continuous $q$-Laguerre generating function (\ref
{genrating1}) as a multiplicative series of the classical Laguerre
generating function, we introduce the following parameters
\begin{eqnarray}
x_k&=&\frac{-q^{(\alpha+\frac{1}{2})k}-q^{(\alpha+1)k}+2q^{(\frac{1}{2}\alpha+\frac{1}{4})k}\cos
k\theta}{k(1-q^k)}\nonumber\\
\tau_k&=&\frac{t^k}{t^k-1}
\end{eqnarray}
then we have
\begin{eqnarray}\label{geni}
\textsf{G}_q^{\alpha}(x,t)&=&\prod_{k\in\mathbb{N}^*}\left[ \textsf{G}^{\alpha_k}(x_k,\tau_k)(1-\tau_k)^{\alpha_k+1}\right]\\
&=&\sum_{\{n_k\}}\prod_{k\in\mathbb{N}^*}\left[L_{n_k}^{\alpha_k}(x_k)\tau_
k^{n_k}(1-\tau _k)^{\alpha_k+1}\right]
\end{eqnarray}
For instance $\{\alpha_k\}$ is a family of generic parameters. We
rewrite
\begin{equation}
\tau_ k^{n_k}(1-\tau _k)^{\alpha_k+1}=(-1)^{n_k}\sum_{m_k\geq
0}\frac{(\alpha_k+n_k+1)_{m_k}}{m_k!}t^{k(n_k+m_k)}
\end{equation}
We obtain
\begin{eqnarray}\label{genrating3}
\textsf{G}_q^{\alpha}(x,t)=\sum_{\{n_k\}}\sum_{\{m_k\}}\prod_{k\in\mathbb{N}^*}\left[(-1)^{n_k}L_{n_k}^{\alpha_k}(x_k)\frac{(\alpha_k+n_k+1)_{m_k}}{m_k!}t^{k(n_k+m_k)}\right]
\end{eqnarray}
Inserting the series given in (\ref{genrating1}) in the later
relation (\ref{genrating3}) and comparing coefficients of equal
power in $t$ on both sides, we obtain our connection formula for the
continuous $q$-Laguerre polynomials in terms of their classical
analogues
\begin{equation}\label{connection}
P_{n}^{\alpha}(x|q)=\sum_{\{n_k\}}\sum_{\{m_k\}}\prod_{k\in\mathbb{N}^*}\left[(-1)^{n_k}L_{n_k}^{\alpha_k}(x_k)\frac{(\alpha_k+n_k+1)_{m_k}}{m_k!}\right]
\delta_{\sum_{k\in\mathbb{N}^*}k(n_k+m_k),n}.
\end{equation}
It's obvious that the family $\{\alpha_k\}$ could be any real
parameters and by construction the left hand side of
(\ref{connection})
 must be independent of this family. Each set $\{\alpha_k\}$
 provides
 an expansion of the continuous $q$-Laguerre polynomials. The solutions of the Diophantine partition relation
\begin{equation}\label{partition}
\sum_{k\in\mathbb{N}^*}k(n_k+m_k)=n
\end{equation}
determine the set of classical Laguerre polynomials contributing to
the expansion of the continuous $q$-Laguerre polynomial. For an
explicit example we have used the connection formula
(\ref{connection}) to write the $P_{4}^{\alpha}(x|q)$ polynomial.
This is done after solving the Diophantine partition equation
(\ref{partition}) for $n=4$. In Table \ref{Table 1} we have listed
the corresponding solutions for this case together with their
respective classical Laguerre polynomials contributions to the
connection formula (\ref{connection}).
 \vskip 0.5cm
\section{Continuous big $q$-Hermite polynomials}
\setcounter{equation}{0}

The continuous big $q$-Hermite polynomials $H_n(x;a;q)$ appear in
many contexts of mathematical physics in particular in \cite {FTV}
where it was shown that they realize a basis for a representation
space of an extended $q$-oscillator algebra. They depend on one
parameter and are defined by \cite {KS}
\begin{eqnarray}
H_n(x;a;q)&=&a^{-n}{}_3\phi_2\left(q^{-n}, ae^{i\theta},ae^{-i\theta};0,0;q;q\right)\nonumber\\
&=& e^{in\theta}{}_2\phi_0\left(q^{-n},
ae^{i\theta};-;q;q^{n}e^{-2i\theta}\right),\quad x=\cos\theta.
\end{eqnarray}
And their generating function is given by
\begin{eqnarray}\label {gen1}
\textsc{G}_q(x,a,t)&=&\frac{(at;q)_{\infty}}{(e^{i\theta t};q)(e^{-i\theta t};q)},\qquad x=\cos\theta\nonumber\\
&=&E_q(-at)e_q(e^{i\theta}t)e_q(e^{-i\theta}t)
=\sum_{n=0}^{\infty}\frac{H_n(x;a;q)}{(q;q)_{n}}t^n.
\end{eqnarray}
Let's recall that the classical Hermite polynomials, defined by
\cite {AAR}
\begin{equation}
H_n(x)=(2x)^n {}_2F_0\left(-n/2,-(n-1)/2;-;-\frac{1}{x^2}\right),
\end{equation}
 can be obtained from the continuous big $q$-Hermite polynomials by the following limit
\begin{eqnarray}
\lim_{q\rightarrow
1}(\frac{1-q}{2})^{-\frac{n}{2}}H_n\left(x(\frac{1-q}{2})^{\frac{1}{2}};a(2(1-q))^{\frac{1}{2}};q\right)=H_n(x-a).
\end{eqnarray}
Their generating function is given by
\begin{eqnarray}\label {gen2}
\textsc{G}(x,t)=\exp(2xt-t^2)=\sum_{n=0}^{\infty}\frac{H_n(x)}{n!}t^n.
\end{eqnarray}
In similar construction, the two kinds of the $q$-exponential
function in the deformed generating function (\ref {gen1}) are
substituted by their expressions given in (\ref{quesne1}) and
(\ref{quesne2}) to obtain the following expression
\begin{equation}
\textsc{G}_q(x,a,t)=\prod_{k\geq1}\exp\left(\frac{-a^k+2\cos(k\theta)}{k(1-q^k)}t^k\right)
\end{equation}
For our purpose we set
\begin{equation}
x_k=\frac{-a^k+2\cos(k\theta)}{2k(1-q^k)}
\end{equation}
then we can write the deformed generating function
$\textsc{G}_q(x,a,t)$ as an infinite product series of the classical
Hermite generating function:
\begin{equation}\label{gen3}
\textsc{G}_q(x,a,t)=\prod_{k\geq1}\left(\textsc{G}(x_k,t^k)e^{t^{2k}}\right)
\end{equation}
Inserting the series given in rhs of (\ref{gen1}) and (\ref{gen2})
and the series of $e^{t^{2k}}$ in relation (\ref{gen3}); and
comparing coefficients of equal power in $t$ on both sides, we
obtain our connection formula for the continuous big $q$-Hermite
polynomials in terms of their classical analogues
\begin{equation}\label{connection2}
 \frac{H_n(x;a;q)}{(q;q)_{n}}=\sum_{\{n_k\}}\sum_{\{m_k\}}\prod_{k\in\mathbb{N}^*}\frac{H_{n_k}(x_k)}{n_k!m_k!}\delta_{\sum_{k\in\mathbb{N}^*}k(n_k+2m_k),n}.
\end{equation}
Here again all the problem stands in finding the solutions of the
Diophantine partition equation
\begin{equation}\label{Diaph1}
\sum_{k\in\mathbb{N}^*}k(n_k+2m_k)=n. \end{equation} To illustrate
the connection formula (\ref{connection2}) we have listed in Table
\ref {Table 2} all the possible non-zero solutions of (\ref{Diaph1})
for $n=5$ and their corresponding classical Hermite polynomials
involved in the construction of $H_5(x;a;q)$.
 {\remark
The continuous $q$-Hermite polynomials, (see, for example,
\cite{KS}, section 3.26), can easily be obtained from the continuous
big $q$-Hermite polynomials $H_n(x;a;q)$ by replacing $a=0$, then we
can derive their connection formula in terms of the classical
Hermite polynomials by taking $a=0$ in both sides of
(\ref{connection2}).}
 \vskip 0.5cm

\section{$q$-Meixner-Pollaczek polynomials}
\setcounter{equation}{0}

In this section we treat the cases of the $q$-Meixner-Pollaczek
polynomials \cite{KS}
\begin{eqnarray}
P_n(x;\lambda;q)&=&q^{-n\lambda}e^{-in\phi}\frac{(q^{2\lambda};q)_n}{(q;q)_n}{}_3\phi_2\left(q^{-n},q^{\lambda}e^{i(\theta+2\phi)}, q^{\lambda}e^{-i\theta};q^{2\lambda},0;q;q\right),\quad x=\cos(\theta+\phi)\nonumber\\
&=&\frac{(q^{\lambda}e^{-i\theta};q)_n}{(q;q)_n}e^{in(\theta+\phi)}{}_2\phi_1\left(q^{-n},q^{\lambda}e^{i\theta};q^{1-\lambda-n}e^{i\theta};q;q^{1-\lambda}e^{-i(\theta+2\phi)}\right)
.
\end{eqnarray}
These are the $q$-analogue of the classical Meixner-Pollaczek
polynomials defined by \cite{AAR}
\begin{eqnarray}
P_n^{(\lambda)}(x;\phi)=\frac{(2\lambda)_n}{n!}e^{in\phi}{}_2F_1\left(-n,\lambda+ix;2\lambda;1-e^{-2i\phi}\right),\quad
\lambda >0,\quad 0<\phi<\pi.
\end{eqnarray}
It's obvious from the last two expressions that the following limit
is true:
\begin{eqnarray}
\lim_{q\rightarrow 1}P_n(\cos(\ln
q^{-x}+\phi);\lambda;q)=P_n^{(\lambda)}(x;-\phi)
\end{eqnarray}
The generating function of the $q$-Meixner-Pollaczek polynomials is
given by
\begin{eqnarray}\label{gen4}
\textbf{G}_q^{\lambda}(x,t)&=&\left|\frac{(q^{\lambda}e^{i\phi}t;q)_{\infty}}{(e^{i(\theta+\phi)}t;q)_{\infty}}\right|^2
=\frac{E_q(-q^{\lambda}e^{i\phi}t)E_q(-q^{\lambda}e^{-i\phi}t)}{E_q(-e^{i(\theta+\phi)}t)E_q(-e^{-i(\theta+\phi)}t)}\\
&=&\sum_{n=0}^{\infty}P_n(x;\lambda;q)t^n,\qquad
x=\cos(\theta+\phi).\nonumber
\end{eqnarray}
Which can be written, after using (\ref{quesne2}), as
\begin{eqnarray}
\textbf{ G}_q^{\lambda}(x,t) &=&
\prod_{k\geq1}\exp\left(\frac{2}{k}\frac{-q^{k\lambda}\cos
k\phi+\cos k(\theta+\phi)}{1-q^k}t^k\right)
\end{eqnarray}
We set
\begin{equation}
    x_k=\frac{2}{k}\frac{-q^{k\lambda}\cos k\phi+\cos k(\theta+\phi)}{1-q^k}
\end{equation}
then we can write $
 \textbf{\textbf{G}}_q^{\lambda}(x,t) $ as
 \begin{equation}\label{GeneratingM}
 \textbf{G}_q^{\lambda}(x,t)
 =\sum_{\{n_k\}}\prod_{k\geq1}\left[\frac{1}{n_k!}x_k^{n_k}t^{kn_k}\right]
\end{equation}
On the other hand, for any $x\in\mathbb{R}$ and $m\in \mathbb{N}$, we
can expand $x^m$ with respect of the classical Meixner-Pollaczek
polynomials in the following way
\begin{equation}\label {expansionM}
x^m=\sum_{l=0}^m A_{l,m}^{\lambda,\phi}P_{l}^{(\lambda)}(x;\phi)
\end{equation}
where the $A_{l,m}^{\lambda,\phi}$ satisfy, for $l=0,...,m$ , the
following recursion relation
\begin{equation}\label{}
\left\{
\begin{array}{l}
A_{0,0}^{\lambda,\phi}=1\\2\sin \phi
A_{l,m+1}^{\lambda,\phi}=(l+2\lambda)A_{l+1,m}^{\lambda,\phi}-2(l+\lambda)\cos\phi
A_{l,m}^{\lambda,\phi}+l A_{l-1,m}^{\lambda,\phi},
\end{array} \right.
\end{equation}
and for $l>m$, $A_{l,m}^{\lambda,\phi}=0$. Using this expansion in
the rhs of (\ref {GeneratingM}) to express $x_k^{n_k}$ in term of
the classical Meixner-Pollaczek polynomials and comparing
coefficients of equal power in $t$ on both sides, we obtain our
connection formula for the $q$-Meixner-Pollaczek polynomials in
terms of their classical partners of lower dimensions:
\begin{equation}\label{connectionM}
P_n^{(\lambda)}(x;\phi;q)=\sum_{n_1,n_2,...=0}^{\infty}\sum_{{{0\leq
l_1\leq n_1,} \atop {0\leq l_2\leq
n_2,}}\atop{\ldots}}\prod_{k\geq1}\left[\frac{1}{n_k!}A_{l_k,n_k}^{\lambda_k,\phi_k
}P_{l_k}^{(\lambda
_k)}(x_k;\phi_k)\right]\delta_{\sum_{k\geq1}kn_k,n}
\end{equation}
Here, the same observation made above for the $\{\alpha_k\}$ family
of the continuous $q$-Laguerre polynomials occurs for the
$\{\lambda_k\}$ and $\{\phi_k\}$ families in (\ref {connectionM})
i.e. the later connection formula remains independent of the
$\lambda_k$ and $\phi_k$ parameters. As example we list in below a
few calculus of $q$-Meixner-Pollaczek polynomials in terms of their
classical counterparts, after solving the partition equation
$\sum_{k\geq1}kn_k=n$ in each cases.

\begin{eqnarray}
  P_0^{(\lambda)}(x;\phi;q)&=&P_0^{(\lambda)}(x;\phi)=1 \nonumber\\
  P_1^{(\lambda)}(x;\phi;q) &=& \frac{1}{2\sin\phi_1}\left[P_1^{(\lambda_1)}(x_1;\phi_1)-2\lambda_1\cos\phi_1 P_0^{(\lambda_1)}(x_1;\phi_1)\right] \nonumber\\
  P_2^{(\lambda)}(x;\phi;q)&=&\frac{1}{4\sin^2\phi_1}\left[P_2^{(\lambda_1)}(x_1;\phi_1)-(2\lambda_1+1)\cos\phi_1P_1^{(\lambda_1)}(x_1;\phi_1)\right.\nonumber\\&+&\left.\lambda_1(2\lambda_1\cos^2\phi_1+1)P_0^{(\lambda_1)}(x_1;\phi_1)\right]\nonumber\\&+&\frac{1}{2\sin\phi_2}\left[P_1^{(\lambda_2)}(x_2;\phi_2)-2\lambda_2\cos\phi_2 P_0^{(\lambda_2)}(x_2;\phi_2)\right]\nonumber \\
  P_3^{(\lambda)}(x;\phi;q)&=&\frac{1}{24\sin^3\phi_1}\left[3P_3^{(\lambda_1)}(x_1;\phi_1)-6(\lambda_1+1)\cos\phi_1P_2^{(\lambda_1)}(x_1;\phi_1)\right.\nonumber\\
  &+&\left.((3\lambda_1+1)+2(3\lambda_1^2+3\lambda_1+1)\cos^2\phi)P_1^{(\lambda_1)}(x_1;\phi_1)\right.\nonumber\\&-&2\left.\lambda_1\cos\phi_1(3\lambda_1+1+2\lambda_1^2\cos^2\phi_1)P_0^{(\lambda_1)}(x_1;\phi_1)\right]\nonumber\\
  &+&\frac{1}{4\sin\phi_1\sin\phi_2}\left[P_1^{(\lambda_1)}(x_1;\phi_1)-2\lambda_1\cos\phi_1
  P_0^{(\lambda_1)}(x_1;\phi_1)\right]\times\nonumber\\
  &\times&\left[P_1^{(\lambda_2)}(x_2;\phi_2)-2\lambda_2\cos\phi_2
  P_0^{(\lambda_2)}(x_2;\phi_2)\right]\nonumber\\
  &+&\frac{1}{2\sin\phi_3}\left[P_1^{(\lambda_3)}(x_3;\phi_3)-2\lambda_3\cos\phi_3
  P_0^{(\lambda_3)}(x_3;\phi_3)\right]
\end{eqnarray}
\vskip 0.5cm

\section{$q$-Gegenbauer polynomials}
\setcounter{equation}{0}

The $q$-Gegenbauer (or continuous $q$-ultraspherical or Rogers)
polynomials are given by \cite{GR}
\begin{eqnarray}
    C_n^{(\lambda)}(x;q)&=&\frac{(q^{2\lambda};q)_n}{(q;q)_{n}}q^{-\frac{n\lambda}{2}}
{}_4\phi_3\left(q^{-n} ,q^{2 \lambda+ n},
q^{\frac{\lambda}{2}}e^{i\theta}, q^{\frac{\lambda}{2}}e^{-i\theta}
;q^{\lambda+\frac{1}{ 2} }, -q^{\lambda},-q^{\lambda+\frac{1}{ 2}
};q;q
\right)\\
&=&\frac{(q^{2\lambda};q)_n}{(q;q)_{n}}q^{-n\lambda}e^{-in\theta}
{}_3\phi_2\left(q^{-n} ,q^{ \lambda}, q^{\lambda}e^{2i\theta}
;q^{2\lambda}, 0;q;q
\right)\nonumber\\
&=&\frac{(q^{\lambda};q)_n}{(q;q)_{n}}e^{in\theta}
{}_2\phi_1\left(q^{-n} ,q^{ \lambda} ;q^{1-n-\lambda},
q;e^{-2i\theta} \right),\qquad x=\cos\theta.\nonumber
\end{eqnarray}
These polynomials can also be written as
\begin{equation}\label{geg}
    C_n^{(\lambda)}(x;q)=\sum_{k=0}^{n}\frac{(q^{\lambda};q)_{k}(q^{\lambda};q)_{n-k}}{(q;q)_{k}(q;q)_{n-k}}e^{i(n-2k)\theta},\qquad x=\cos\theta.
\end{equation}
which are the $q$-analogues of the classical Gegenbauer (or
ultraspherical) polynomials \cite{AAR}
\begin{eqnarray}
    C_n^{(\lambda)}(x)&=&\frac{(2\lambda)_n}{n!}{}_2F_1\left(-n,n+2\lambda;\lambda+\frac{1}{2};\frac{1-x}{2}\right),\quad
    \lambda\neq0\\
    &=&\sum_{k=0}^{n}\frac{({\lambda})_{k}({\lambda})_{n-k}}{k!(n-k)!}e^{i(n-2k)\theta},\qquad x=\cos\theta.
\end{eqnarray}
The generating function of the $q$-Gegenbauer polynomials is given
by
\begin{eqnarray}\label {generating8}
\verb"G"_q^{\lambda}(x;t)&\equiv
&\frac{(q^{\lambda}e^{i\theta}t;q)_\infty(q^{\lambda}e^{-i\theta}t;q)_\infty}{(e^{i\theta}t;q)_\infty(e^{-i\theta}t;q)_\infty},\qquad
    x=\cos\theta.\nonumber\\
&=&\frac{E_q(-q^{\lambda}e^{i\theta}t)E_q(-q^{\lambda}e^{-i\theta}t)}{E_q(-e^{i\theta}t)E_q(-e^{-i\theta}t)}
=\sum_{n\geq 0}C_n^{(\lambda)}(x;q)t^n
\end{eqnarray}
Again by the mean of (\ref{quesne2}) the last generating function
(\ref{generating8}) take the following form
\begin{eqnarray}
\verb" G"_q^{\lambda}(x,t) &=&
\exp\left(\sum_{k\geq1}\frac{2}{k}\frac{1-q^{k\lambda}}{1-q^k}\cos
 (k\theta)
 t^k \right)
\end{eqnarray}
With the parametrization $ x_k=\cos(k\theta)$ and $
[\lambda]_{q^k}=\frac{1-q^{k\lambda}}{1-q^k}$ the deformed
generating function reads
\begin{equation}\label{generating10}
\verb"G"_q^{\lambda}(x,t)=\sum_{n_k\geq0}\prod_{k\geq1}\left[\frac{1}{n_k!}\left(\frac{2}{k}[\lambda]_{q^k}\right)^{n_k}x_k^{n_k}t^{n_k}\right].
\end{equation}
Recall that for any $|x|<1$ and $m\in \mathbb{N}$, we can expand
$x^{m}$ with respect of the classical Gegenbauer polynomials as
  \begin{equation}\label {expansion}
x^m=\frac{m!}{2^m}\sum_{l=0}^{[\frac{m}{2}]}a_{l,m}^{\lambda}C_{2l+s}^{(\lambda)}(x)
\end{equation}
with
  \begin{equation}
a_{l,m}^{\lambda}=\frac{\Gamma(\lambda)(2l+s+\lambda)}{\Gamma([\frac{m}{2}]+l+s\lambda+1)([\frac{m}{2}]-l)!}
\end{equation}
and $s=0$ ( resp. $s=1$) for $m$ even ( resp. $m$ odd).
$[\frac{m}{2}]$ denotes the largest integer smaller than or equal to
$\frac{m}{2}$. Using this expansion in the rhs of (\ref
{generating10}) to rewrite $x_k^{n_k}$ in terms of the classical
Gegenbauer polynomials, and comparing coefficients of equal power in
$t$ on both sides, we obtain a new connection formula for the
$q$-Gegenbauer polynomials in terms of their classical analogues,
which is more general than the one found in \cite {CJN}

\begin{equation}\label{connection4}
C_n^{(\lambda)}(x;q)=\sum_{n_1,n_2,...=0}^{\infty}\sum_{{{0\leq
l_1\leq n_1,}\atop{0\leq l_2\leq
n_2,}}\atop{\ldots}}\prod_{k\geq1}\left[\left(\frac{1}{k}[\lambda]_{q^k}\right)^{n_k}a_{l_k,n_k}^{\lambda
_k}C_{2l_k+s_k}^{(\lambda
_k)}(x_k)\right]\delta_{\sum_{k\geq1}kn_k,n}
\end{equation}

Here also the connection formula (\ref {connection4}) remains
independent of the $\lambda_k$ parameters. Each real set
$\{\lambda_k\}$ provides an expansion of the $q$-Gegenbauer
polynomials. As an illustration we here list the calculus of the
first six $q$-Gegenbauer polynomials in terms of their classical
counterparts.

\begin{eqnarray}
C_0^{(\lambda)}(x;q)&=& C_0^{(\lambda)}(x)=1. \nonumber\\
  C_1^{(\lambda)}(x;q) &=& [\lambda]_{q}\frac{1}{\lambda_1} C_1^{(\lambda_1)}(x_1).\nonumber\\
C_2^{(\lambda)}(x;q) &=& [\lambda]_{q}^2\frac{1}{\lambda_1+1}\left( C_0^{(\lambda_1)}(x_1)+\frac{1}{\lambda_1}C_2^{(\lambda_1)}(x_1)\right)+[\lambda]_{q^2}\frac{1}{2\lambda_2}C_1^{(\lambda_2)}(x_2).\nonumber\\
 C_3^{(\lambda)}(x;q) &=&[\lambda]_{q}^3\frac{1}{\lambda_1+2}\left(\frac{1}{\lambda_1}C_1^{(\lambda_1)}(x_1)+\frac{1}{\lambda_1(\lambda_1+1)}C_3^{(\lambda_1)}(x_1)\right)\nonumber \\
 &+&\frac{1}{2}[\lambda]_{q}[\lambda]_{q^2}\frac{1}{\lambda_1\lambda_2}C_1^{(\lambda_1)}(x_1)C_1^{(\lambda_2)}(x_2)
 +\frac{1}{3}[\lambda]_{q^3}\frac{1}{\lambda_3}C_1^{(\lambda_3)}(x_3).
 \nonumber\\
  C_4^{(\lambda)}(x;q)&=&[\lambda]_{q}^4\frac{1}{\lambda_1+1}\left(\frac{1}{(2\lambda_1+2)}C_0^{(\lambda_1)}(x_1)+\frac{1}{\lambda_1(\lambda_1+3)}C_2^{(\lambda_1)}(x_1)\right.\nonumber\\
  &+&\left.\frac{1}{\lambda_1(\lambda_1+2)(\lambda_1+3)}C_4^{(\lambda_1)}(x_1)\right)\nonumber \\
  &+&\frac{1}{2}[\lambda]_{q^2}[\lambda]_{q}^2\frac{1}{\lambda_2(\lambda_1+1)}\left(C_0^{(\lambda_1)}(x_1)+\frac{1}{\lambda_1}C_2^{(\lambda_1)}(x_1)\right)C_1^{(\lambda_2)}(x_2)\nonumber \\
  &+&
  \frac{1}{3}[\lambda]_{q}[\lambda]_{q^3}\frac{1}{\lambda_1\lambda_3}C_1^{(\lambda_1)}(x_1)C_1^{(\lambda_3)}(x_3)+\frac{1}{4}[\lambda]_{q^4}\frac{1}{\lambda_4}C_1^{(\lambda_4)}(x_4).\nonumber
\end{eqnarray}
\begin{eqnarray}
C_5^{(\lambda)}(x;q)&=&[\lambda]_{q}^5\frac{1}{\lambda_1(\lambda_1+2)}\left(\frac{1}{2(\lambda_1+3)}C_1^{(\lambda_1)}(x_1)+\frac{1}{(\lambda_1+1)(\lambda_1+4)}C_3^{(\lambda_1)}(x_1)\right.\nonumber\\&+&\left.\frac{1}{(\lambda_1+1)(\lambda_1+3)(\lambda_1+4)}C_5^{(\lambda_1)}(x_1)\right)
\nonumber\\&+&\frac{1}{2}[\lambda]_{q}^3[\lambda]_{q^2}\frac{1}{\lambda_1\lambda_2(\lambda_1+2)}\left(C_0^{(\lambda_1)}(x_1)+\frac{1}{\lambda_1+1}C_3^{(\lambda_1)}(x_1)\right)C_1^{(\lambda_2)}(x_2)\nonumber
\\&+&\frac{1}{3}[\lambda]_{q}^2[\lambda]_{q^3}\frac{1}{\lambda_3(\lambda_1+1)}\left(C_0^{(\lambda_1)}(x_1)+\frac{1}{\lambda_1}C_2^{(\lambda_1)}(x_1)\right)C_1^{(\lambda_3)}(x_3)\nonumber\\
&+&[\lambda]_{q}[\lambda]_{q^4}\frac{1}{\lambda_1\lambda_4}C_1^{(\lambda_1)}(x_1)C_1^{(\lambda_4)}(x_4)+\frac{1}{6}[\lambda]_{q^2}[\lambda]_{q^3}\frac{1}{\lambda_2\lambda_3}C_1^{(\lambda_2)}(x_2)C_1^{(\lambda_3)}(x_3)\nonumber\\
&+&\frac{1}{4}[\lambda]_{q}[\lambda]_{q^2}^2\frac{1}{\lambda_1(\lambda_2+1)}\left(C_0^{(\lambda_2)}(x_2)+\frac{1}{\lambda_2}C_2^{(\lambda_2)}(x_2)\right)C_1^{(\lambda_1)}(x_1)+\frac{1}{5}[\lambda]_{q^5}\frac{1}{\lambda_5}C_1^{(\lambda_5)}(x_5).\nonumber\\
\end{eqnarray}

{\remark

Recall that the continuous $q$-Legendre polynomials, denoted
$P_n(x|q)$ (see \cite{KS} {subsection 3.10.2} ), are related to the
$q$-Gegenbauer polynomials by
\begin{equation}
P_n(x|q)=q^{\frac{n}{4}}C^{(\frac{1}{2})}_n(x;q)
\end{equation}and the classical Legendre polynomials can be obtained
from the classical Gegenbauer polynomials by replacing
$\lambda=\frac{1}{2}$, then we can derive a connection formula
between continuous $q$-Legendre and classical Legendre polynomials
by taking $\lambda=\lambda_k=\frac{1}{2}$, $\forall k\in
\mathbb{N}^*$ in (\ref {connection4}), and multiplying both sides by
$q^{\frac{n}{4}}$.}
 \vskip 0.5cm
\section{Conclusion and discussion}
In this work, we had successfully written the connection formulae of
some $q$-orthogonal polynomials appearing in the Askey scheme
\cite{KS}. The first one was the continuous $q$-Laguerre
polynomials, which are representing others $q$-analogues of the
classical Laguerre polynomials. An explicit example $
P_4^{\alpha}(x| q)$ was given. It follows from these results that
the solutions of the Diophantine equation fix the finite dependence
structure between classical polynomials $ L_n^{\alpha}(x)$ and
deformed polynomials $ P_n^{\alpha}(x| q)$ for any fixed $n$. The
obtention of the connection formulae was possible only because the
generating function of continuous $q$-Laguerre polynomials are the
product of Jackson's $q$-exponentials which could be expressed in
more useful forms found by C. Quesne \cite{Q}. Our second sample in
the Askey scheme was the continuous big $q$-Hermite. In this case,
we had used the same arguments and method as in the precedent
example and our connection formula was supported by an explicit
example. The third polynomials in the list of this work were the
$q$-Meixner-Pollaczek ones. The uses of relations (\ref{quesne1})
and (\ref{quesne2}) obtained by \cite {Q} and series expansion
allowed us to write a well defined connection formula relating the
deformed polynomials to their classical counterparts. Several
examples were given. In the last section, it wasn't difficult to
give the connection formula of
 $q$-Gegenbauer polynomials in more general form than the one given in \cite{CJN}.
In all cases, except in the big $q$-Hermite polynomials cases, the
generic family parameters which appear in computation process, drop
out by construction in the final results. This means that quantum
deformation of such orthogonal polynomials is not bijective.
However, for the others polynomials in the Askey scheme possessing
generating functions not expressed in product of $q$-exponentials,
the above prescription stops working. The case of Bessel functions,
which are not orthogonal polynomials, could be a good candidate to
write connection formulae since their generating function uses
Jackson's $q$-exponentials; but this is not an easy task because the
derived Diophantine partition equation couldn't be solved so easily,
even in the simplest cases. However, our results may be useful in
finding the relations of matrix elements of the unitary
co-representations of the some quantum group associated with
$q$-orthogonal polynomials.

For instance, we can notice  from \cite{FV} the
existence of simple relations between the matrix elements of the
metaplectic representation of $su_q(1,1)$ and $q$-generalization of the Gegenbauer
polynomials which are slightly different from (\ref{geg}) i.e. the expressions (22) and (23). For these polynomials, we can
easily compute their associated connection formula and
use it, after setting $\lambda=-(n+m), -(n+m+1)$ and
$\lambda_k=-(n_k+m_k), -(n_k+m_k+1)$, to get a new relation which links the
quantum matrix elements associated with $SU_q(1,1)$ to their classical analogous;
constituting then an infinite dimensional representation of
$SU_q(1,1)$. In some way, this relation may be viewed as a kind of
realization map of the standard deformation of the group from its
non deformed form. But what is more interesting is when one takes a
generic value of $\lambda$; this provides a new continuous
representation more general than the precedent ones.
Whether this representation fits or not with actual known
representations of $SU(1,1)$ is now the question.

\vskip 0.5cm \noindent{\bf Acknowledgements:} {\em
 We would like to thank Prof. B. Abdesselam, L. Frappat and P. Sorba for precious help and useful discussions. We are also grateful to Prof. P. Aurenche for his hospitality in LAPTH (Annecy).
This work is supported by the CMEP program 04 MDU 615. }

\newpage
\begin{table}
\begin{center}
\begin{tabular}{|c||c|}
\hline
Solutions of (\ref{partition}) for $n =4$&Contributions to $P_{4}^{\alpha}(x|q)$ \\
\hline \hline
$n_1=4$ & $L_4^{\alpha_1}(x_1)$ \\
\hline
$m_1=4$ & $\frac{(\alpha_1+1)_4}{4!}$ \\
\hline
$n_1=3$, $m_1=1$ & $-(\alpha_1+4)L_3^{\alpha_1}(x_1)$  \\
\hline
$n_1=1$, $m_1=3$ & $-\frac{(\alpha_1+2)_3}{3!}L_1^{\alpha_1}(x_1)$  \\
\hline
$n_1=2$, $m_1=2$ & $\frac{(\alpha_1+3)_2}{2}L_2^{\alpha_1}(x_1)$   \\
\hline
$n_1=1,m_1=1,n_2=1$ & $(\alpha_1+2)L_1^{\alpha_1}(x_1)L_1^{\alpha_2}(x_2)$  \\
\hline
$n_1=1,m_1=1,m_2=1$ & $-(\alpha_1+2)(\alpha_2+1)L_1^{\alpha_1}(x_1)$  \\
\hline
$n_1=2,n_2=1$ & $-L_2^{\alpha_1}(x_1)L_1^{\alpha_2}(x_2)$  \\
\hline
$n_1=2,m_2=1$ & $(\alpha_2+1)L_2^{\alpha_1}(x_1)$  \\
\hline
$m_1=2,n_2=1$ & $-\frac{(\alpha_1+1)_2}{2}L_1^{\alpha_2}(x_2)$  \\
\hline
$m_1=2,m_2=1$ & $\frac{1}{2}(\alpha_1+1)_2(\alpha_2+1)$  \\
\hline
$n_2=2$ & $L_2^{\alpha_2}(x_2)$ \\
\hline
$m_2=2$ & $\frac{(\alpha_2+1)_2}{2}$  \\
\hline
$n_2=1$, $m_2=1$ & $-(\alpha_2+2)L_1^{\alpha_2}(x_2)$  \\
\hline
$n_1=1, n_3=1$ & $L_1^{\alpha_1}(x_1)L_1^{\alpha_3}(x_3)$  \\
\hline
$m_1=1,n_3=1$ & $-(\alpha_1+1)L_1^{\alpha_3}(x_3)$  \\
\hline
$n_1=1,m_3=1$ & $-(\alpha_3+1)L_1^{\alpha_1}(x_1)$  \\
\hline
$m_1=1,m_3=1$ & $(\alpha_1+1)(\alpha_3+1)$  \\
\hline
$n_4=1$ & $-L_1^{\alpha_4}(x_4)$  \\
\hline
$m_4=1$ & $(\alpha_4+1)$  \\
\hline

\end{tabular}
\caption[smallcaption]{Contributions to $P_{4}^{\alpha}(x|q)$. }
\label{Table 1}
\end{center}
\end{table}
\begin{table}
\begin{center}
\begin{tabular}{|c||c|}
\hline
Solutions of (\ref{Diaph1}) for $n =5$&Contributions to $H_{5}(x,a;q)$ \\
\hline \hline
$n_1=5$ & $\frac{(q;q)_5}{5!}H_5(x_1) $ \\
\hline
$n_1=1$, $m_1=2$ & $\frac{(q;q)_5}{2}H_1(x_1)$ \\
\hline
$n_1=1$, $n_2=2$ & $\frac{(q;q)_5}{2}H_1(x_1)H_2(x_2)$  \\
\hline
$n_1=1$, $n_2=1$, $m_1=1$ & $(q;q)_5H_1(x_1)H_1(x_2)$  \\
\hline
$n_1=2$, $n_3=1$ & $\frac{(q;q)_5}{2}H_1(x_3)H_2(x_1)$   \\
\hline
$n_1=1$, $m_2=1$ & $(q;q)_5H_1(x_1)$ \\
\hline
$n_3=1$, $m_1=1$ & $(q;q)_5H_1(x_3)$  \\
\hline
$n_2=1$, $n_3=1$ & $(q;q)_5H_1(x_2)H_1(x_3)$  \\
\hline
$n_1=1$, $n_4=1$ & $(q;q)_5H_1(x_1)H_1(x_4)$  \\
\hline
$n_5=1$ & $(q;q)_5H_1(x_5)$  \\
\hline
\end{tabular}
\caption[smallcaption]{Contributions to $H_{5}(x,a;q)$. }
\label{Table 2}
\end{center}
\end{table}

\end{document}